\documentclass[a4paper,11pt]{article}
\usepackage{pos}
\renewcommand{\logo}{\relax}

\newcommand{\tr}{\operatorname{tr}}

\newcommand{\dif}{\mathrm{d}}

\newcommand{\pder}[2]{\frac{\partial #1}{\partial #2}}
\newcommand{\iDelta}{\mathit{\Delta}}

\title{Path integral for the closed superstring and the matrix model}

\author*[a,b]{Yuhma Asano}

\affiliation[a]{Institute of Pure and Applied Sciences, University of Tsukuba,\\
1-1-1 Tennodai, Tsukuba, Ibaraki 305-8571, Japan}

\affiliation[b]{Tomonaga Center for the History of the Universe, University of Tsukuba,\\
1-1-1 Tennodai, Tsukuba, Ibaraki 305-8571, Japan}

\emailAdd{asano@het.ph.tsukuba.ac.jp}

\abstract{The IKKT matrix model, which is proposed as a non-perturbative formulation of superstring theory, 
has an issue typical of zero-dimensional theory---ambiguity in 
the definition of its path integral. 
To tackle this issue, 
we revisit the path-integral formulation of perturbative string theory. 
In this article, we review recent progress in the string world-sheet path-integral formulation,
especially in the Minkowski signature.
We first derive the Minkowskian path integral 
of the Nambu-Goto type
equivalent to Polyakov's Euclidean path integral for critical closed string theory,
showing equivalences among 
the Nambu-Goto-, Schild- and Polyakov-type formulations
both in the Minkowskian and Euclidean signatures.
We also show that ``stringy causality'' is realised
in the path-integral formulation at the level of string perturbation theory.
We then obtain the matrix model with a property like the stringy causality,
which turns out to be a Minkowskian version of the NBI-type IKKT matrix model, 
by matrix regularisation of the path integral for perturbative type IIB string theory.}

\FullConference{Proceedings of the Corfu Summer Institute 2025 "School and Workshops on Elementary Particle Physics and Gravity" (CORFU2025)\\
27 April - 28 September, 2025\\
Corfu, Greece\\}

\begin{document}
\maketitle

\section{Introduction}

A starting point of string theory is usually 
a generalisation of the relativistic point-particle theory
to the relativistic string, 
which is formulated by the Nambu-Goto action.
Since we are interested in realistic theory,
we would first consider the Nambu-Goto action in the Minkowski signature.
However, at some point, we somehow move to 
the Polyakov action in the Euclidean signature for convenience
and formulate quantum mechanics of string theory
by Polyakov's Euclidean path integral \cite{Polyakov:1981rd},
which expresses an $n$-point scattering amplitude like
\begin{equation*}
 \mathcal{A}_{j_1,\cdots,j_n}(k_1,\cdots,k_n)
 =\sum_{\chi=2,0,-2,\cdots}g_s^{-\chi}\int\mathcal{D}X \mathcal{D}g\,
 \mathcal{V}_{j_1}(k_1)\cdots \mathcal{V}_{j_n}(k_n)\,
 \exp[-S_{\rm P}^{\rm (E)}]
 ,
\end{equation*}
where $S_{\rm P}^{\rm (E)}$ is the Euclidean Polyakov action.
Although the path-integral formalism is established as 
a standard formulation of string theory,
it is merely a perturbative expansion in the string coupling,
and thus the formulation is not well-defined non-perturbatively.

One of the proposed non-perturbative formulations---the IKKT matrix model, also known as the IIB matrix model---can be 
obtained by matrix regularisation of the type-IIB string world-sheet theory \cite{Ishibashi:1996xs}.
The IKKT matrix model 
has been studied extensively to show the conjecture
that it should describe type IIB string theory
in a non-perturbative manner 
in the limit as the matrix size $N$ approaches infinity
(recent studies of the matrix model include Refs.~\cite{Asano:2024def,Asano:2024edo,Brandenberger:2024ddi,Hartnoll:2024csr,Hattori:2024btt,Komatsu:2024bop,Komatsu:2024ydh,Chou:2025moy,Ciceri:2025maa,Gohara:2025zfh,Hartnoll:2025ecj,Blair:2025nno,Chou:2025rwy,Ho:2025htr,Manta:2025tcl,Laurenzano:2025vfh,Liao:2025yfb,Steinacker:2026qzk,Anagnostopoulos:2026qvz}).
However, there is an ambiguity in definition of the matrix model.
Since the matrix model is zero-dimensional,
there is no canonical formalism in its quantisation,
and thus the matrix model should be defined by path-integral formalism,
which lacks a definitive first principle in zero dimension.
Then, 
since we have a choice from two types of path-integral weight 
and another choice from Euclidean ($G_{\mu\nu}=\delta_{\mu\nu}$) and Minkowski ($G_{\mu\nu}=\eta_{\mu\nu}$) metrics in the matrix-model action 
\begin{equation}
 S(X,\psi;G_{\mu\nu})=\tr(\tfrac{1}{4}G_{\mu\rho}G_{\nu\sigma}[X^\mu,X^\nu][X^\rho,X^\sigma]+\tfrac{1}{2}\psi^TG_{\mu\nu}\Gamma^\mu[X^\nu,\psi]),
\end{equation}
there are basically four types of definition:
\begin{table}[htbp]
 \centering
 \begin{tabular}{c|c|c|c}
  \hline\hline
  \multicolumn{2}{c|}{}&\multicolumn{2}{c}{metric in the action}\\
  \hline
  \multicolumn{2}{c|}{}&Euclidean&Minkowski\\
  \hline
  weight&Euclidean&$e^{-S(X,\psi;\delta_{\mu\nu})}$&$e^{-S(X,\psi;\eta_{\mu\nu})}$\\
  \cline{2-4}
  &Minkowski&$e^{iS(X,\psi;\delta_{\mu\nu})}$&$e^{iS(X,\psi;\eta_{\mu\nu})}$\\
  \hline\hline
 \end{tabular}
 \label{tab:}
\end{table}%

\noindent
Among them,
we usually do not take $e^{-S(X,\psi;\eta_{\mu\nu})}$ 
as a path-integral weight
because the partition function is not convergent on its own. 
In addition,
there is another type of ambiguity
in path integrals with weights of the Minkowski type, $e^{iS}$,
because they are conditionally convergent at best
and depend on the prescription for the convergence.
Note that 
the IKKT models with the weight $e^{-S(X,\psi;\delta_{\mu\nu})}$,
which is conventionally called the ``Euclidean IKKT model'',
the weight $e^{iS(X,\psi;\eta_{\mu\nu})}$,
which is called the ``Minkowskian (Lorentzian) IKKT model'',
and the weight $e^{iS(X,\psi;\delta_{\mu\nu})}$
are not equivalent in general
but these three theories can become equivalent 
if the Minkowskian versions are defined
by a Lorentz-symmetry breaking regulator \cite{Asano:2024def,Asano:2024edo}.

Since the path-integral formulation obtained by matrix regularisation
depends on the form of the original world-sheet action,
we may expect understanding of the path-integral formalism of string world-sheet theory is essential for its non-perturbative formulation by matrices.
It then leads us to the following fundamental questions:
\begin{itemize}
 \item Is Polyakov's Euclidean path integral quantum-mechanically
 equivalent to a theory with the Nambu-Goto action in the Minkowski signature?
 \item The Minkowskian Nambu-Goto action has an ambiguity of sign when the world-sheet area is space-like.
 What is a correct or sensible definition of 
 path-integral formulation of the Minkowskian Nambu-Goto string?
\end{itemize}

\begin{figure}[htbp]
 \centering
 \includegraphics[scale=0.6]{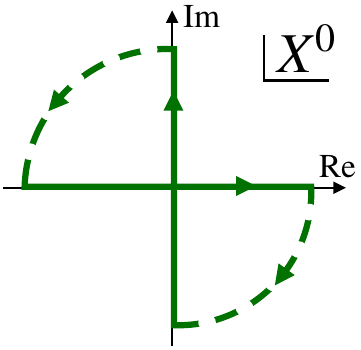}
 \caption{The contour of integration over $X^0$.
  The integration on the closed contour consisting of the solid lines together with the dashed curves is zero because of Cauchy's integral theorem. The vanishing contributions from the arcs (dashed curves) equate the integral on the real axis and that on the imaginary axis in the opposite direction.}
 \label{fig:X0-contour}
\end{figure}

Although the first question may look trivially positive,
it is actually not trivial at all
because the first step, equivalence between Polyakov's Euclidean path integral
and the path integral with the Minkowskian Polyakov action,
is already far non-trivial.
In order to illustrate the non-triviality,
let us consider a bosonic path integral with the Minkowskian Polyakov action,
\begin{equation*}
 \int\mathcal{D}X \mathcal{D}\gamma\,
 \exp\left[
  -i\int\dif\sigma^0\dif\sigma^1\sqrt{-\gamma}\gamma^{ab}\partial_aX^\mu\partial_bX_\mu
 \right]
 ,
\end{equation*}
where $a,b$ take values from 0 and 1, $\mu,\nu$ run from 0 to $D-1$ and
$\gamma^{ab}=e_\alpha^{\;\, a}\eta^{\alpha\beta}e_\beta^{\;\, b}$ is the Lorentzian world-sheet metric 
with $\alpha,\beta=0,1$.
We denote the determinant of the matrix $\gamma_{ab}$ by $\gamma$.
By naively Wick-rotating it by
$X^0=e^{-i\theta}X^D$, $e_0^{\;\, a}=e^{i\theta}e_2^{\;\, a}$
with $\theta$ varying from 0 to $\pi/2$,
and replacing $\sigma^0\to\sigma^2$,
we formally rewrite the path integral as
\begin{align*}
 \int\mathcal{D}X \mathcal{D}\gamma\,
 \exp\bigg[
  -i\int\dif\sigma^1\dif\sigma^2\sqrt{e^2}\bigg(
 &e^{-i\theta}e_1^{\;\, a}e_1^{\;\, b}\partial_aX^i\partial_bX^i
 -e^{i\theta}e_2^{\;\, a}e_2^{\;\, b}\partial_aX^i\partial_bX^i
 \nonumber \\
 &-e^{-3i\theta}e_1^{\;\, a}e_1^{\;\, b}\partial_aX^D\partial_bX^D
 +e^{-i\theta}e_2^{\;\, a}e_2^{\;\, b}\partial_aX^D\partial_bX^D
 \bigg)
 \bigg]
 .
\end{align*}
If the real part of the exponent is negative definite
during the rotation, during which $\theta$ varies from 0 to $\pi/2$,
one can conclude that the original path integral and the Wick-rotated one
are equivalent because a path integral on the closed contour of $X^0(\sigma)$ 
depicted in Fig.~\ref{fig:X0-contour}
vanishes thanks to Cauchy's integral theorem
and the contributions from the arcs vanish as well,
which results in equating the two path integrals.
However,
the real part of the third term in the exponent is positive definite
when $\theta\in(0,\frac{\pi}{3})$,
and thus the naive Wick rotation does not guarantee
the equivalence between the path integrals
with the Euclidean Polyakov action and its Minkowskian version.

The quantum-mechanical equivalence between the Nambu-Goto action and the Polyakov action is also non-trivial.
At the classical level, 
it is straightforward to show the equivalence of these actions,
together with the Schild-type action.
On the other hand, at the path-integral level,
one needs to take care of the measure of path integration.
In the Euclidean case,
the quantum-mechanical equivalence with the Nambu-Goto-type formulation 
was discussed 
in Ref.~\cite{Polyakov:1987ez} for the Polyakov type,
and in Ref.~\cite{Yoneya:1997gs} for the Schild type.
The quantum-mechanical equivalence including the Minkowskian case
has not been completed until Ref.~\cite{Asano:2024edo}.

In this paper,
we will see that 
there is a path-integral formulation of the Minkowskian Nambu-Goto string
equivalent to Polyakov's Euclidean path integral
and that it exhibits ``stringy causality'',
by which a string does not propagate between points at space-like separation.
Moreover,
we will observe that a ``causal'' matrix model is obtained
by the application of matrix regularisation that respects the stringy causality.
We call this matrix model the Minkowskian NBI-type IKKT matrix model
and argue that it can be regarded as a theory of ``causal fuzzy world-sheet''.
This paper is mainly a review of Ref.~\cite{Asano:2024edo} with some updates.

\section{Path integral for the closed string}
Let us start with Polyakov's Euclidean path integral,
then derive equivalent path-integral formulations with different actions,
and finally arrive at the formulation for Min\-kow\-skian Nambu-Goto-type action.
In this paper, we consider
the case of critical bosonic string theory on the flat target space for simplicity.
The equivalences hold true not only for critical bosonic string theory
but also for critical type IIB and IIA string theories on the flat target space.
See Ref.~\cite{Asano:2024edo} for detailed discussions.

\subsection{Euclidean to Minkowskian}\label{sec:EtoM}
Our starting point is the path-integral formulation 
for the Euclidean Polyakov-type action:
\begin{align}
 Z=\int\mathcal{D}X \mathcal{D}g\,
 \exp\left[
 -\frac{1}{2}\int\dif^2\sigma\sqrt{g}g^{ab}h_{ab}
 \right]
 ,
 \label{Polyakov-Euc}
\end{align}
where 
$h_{ab}=\partial_aX^\mu\partial_bX_\mu$ is the induced metric 
with $\mu=1,\cdots,D$ and
$g_{ab}$ is the Riemannian world-sheet metric with $a,b=1,2$.
Note the following arguments and proofs hold true 
even if one inserts operators to the path integral
as long as they have no poles in terms of the dynamical variables.

The world-sheet metric is parameterised by three parameters as
\begin{align}
 g^{ab}=e^{-\phi}
 \begin{pmatrix}
  \frac{(\Lambda^1)^2+(\Lambda^2)^2}{\Lambda^2}&-\frac{\Lambda^1}{\Lambda^2}\\
  -\frac{\Lambda^1}{\Lambda^2}&\frac{1}{\Lambda^2}
 \end{pmatrix}
 ,
\end{align}
where $\phi$ is the conformal factor and $\Lambda^a$ ($a=1,2$) are Lagrange multipliers for the reparametrisation symmetry (diffeomorphism).
We define the measure of integration as
\begin{align}
 \mathcal{D}g
 =\mathcal{D}\phi\,\prod_{\sigma}\frac{2e^\phi\dif\Lambda^1\dif\Lambda^2}{(\Lambda^2)^2}
 ,
\end{align}
induced by the following reparametrisation-invariant norm
\begin{align}
 |\!| \delta g |\!|^2
 &:=
 \frac{1}{2}\int \dif^2\sigma \sqrt{g}\, 
 \left( g^{da} g^{bc} + cg^{ab}g^{cd} \right)
 \delta g_{ab}\, \delta g_{cd}
  \nonumber \\
 &=\int \dif^2\sigma\, e^{\phi}
 \left(
 \frac{(\delta\Lambda^2)^2+(\delta\Lambda^1)^2}{(\Lambda^2)^2}
 +(1+2c)\delta\phi^2
 \right)
 ,
\end{align}
where $c$ is a constant, 
which has no physical effect.

Although the integration over the conformal factor $\phi$ is the source of
the conformal anomaly in general,
the integration becomes trivial in the case of critical string theory ($D=26$ for the bosonic string in the flat space).
Therefore,
since the exponent of the path-integral weight is rewritten as
\begin{align}
 -\frac{1}{2}\int\dif^2\sigma \sqrt{g}g^{ab}h_{ab}=
 -\frac{1}{2}\int\dif^2\sigma\left\{
 \frac{h_{11}}{\Lambda^2}\left(\Lambda^1-\frac{h_{12}}{h_{11}}\right)^2
 +\frac{h_{11}h_{22}-h_{12}^2}{\Lambda^2h_{11}}
 +\Lambda^2h_{11}
 \right\}
 ,
\end{align}
the path integral becomes
\begin{align}
 Z=\int\mathcal{D}X
 \left[ \prod_\sigma \int_0^\infty\frac{2\dif e_g}{(e_g)^{3/2}}
 \right]
 \exp\left[
 -\frac{1}{2}\int\dif^2\sigma\left(
 \frac{h}{e_g}
 +e_g
 \right)
 \right]
 ,
 \label{Schild-Euc}
\end{align}
with the change of variables via $\Lambda^2=e_g/h_{11}$,
after $\phi$ and $\Lambda^1$ are integrated out.
Here, $h$ denotes the determinant of the induced metric $h_{ab}$.
This exponent is the Schild-type action which is
classically equivalent to the Polyakov-type action.
Hence, we find out that the path-integral formulation 
for the Euclidean Schild-type action
with this specific integration measure in Eq.~(\ref{Schild-Euc})
is equivalent to Polyakov's Euclidean path-integral formulation (\ref{Polyakov-Euc}).
Further integration over $e_g$ brings it back to 
the path-integral formulation for the Euclidean Nambu-Goto action.

We now consider equivalence between Euclidean and Minkowskian path-integral formulations.
As seen in Introduction,
the naive Wick rotation in the Polyakov action does not equate
the Euclidean and Minkowskian theories quantum-mechanically.
In contrast, 
let us here apply Wick rotation to the Schild-type path-integral formulation.
We set the deformation of the contour as follows:
\begin{equation}
 X^D=e^{i\theta}X^0
 ,
 \qquad
 e_g=e^{i\theta}e_\gamma
 ,
 \label{Wick-rot-Schild-1}
\end{equation}
with $\theta$ varying from 0 to $\pi/2$,
replacing $\sigma^2\to\sigma^0$.
Here, $e_\gamma$ takes values in the interval $(0,\infty)$,
which is the same as $e_g$.
Then, since $h$ is rewritten as
\begin{align}
 h=\frac{1}{2}\left\{
 (\varepsilon^{ab}\partial_aX^i\partial_bX^j)^2
 +2(\varepsilon^{ab}\partial_aX^D\partial_bX^i)^2
 \right\}
 ,
\end{align}
where $i,j=1,\cdots,D-1$,
the exponent in the path integral is rotated as
\begin{align}
 &-\frac{1}{2}\int\dif^2\sigma\left(
 \frac{1}{2e_\gamma}
 \left\{
 e^{-i\theta}(\varepsilon^{ab}\partial_aX^i\partial_bX^j)^2
 +2e^{i\theta}(\varepsilon^{ab}\partial_aX^0\partial_bX^i)^2
 \right\}
 +e^{i\theta}e_\gamma
 \right)
  \nonumber \\
 &\to -\frac{i}{2}\int\dif^2\sigma\left(
 \frac{1}{2e_\gamma}
 \left\{
 -(\varepsilon^{ab}\partial_aX^i\partial_bX^j)^2
 +2(\varepsilon^{ab}\partial_aX^0\partial_bX^i)^2
 \right\}
 +e_\gamma
 \right)
  ,
  \label{exponent-Wick-rot}
\end{align}
as $\theta$ approaches $\pi/2$.
In this case, 
the real part of the exponent is negative definite
during the rotation.
Therefore, using Cauchy's integral theorem,
we find that the Schild-type formulation (\ref{Schild-Euc})
becomes
\begin{align}
 Z=\int\mathcal{D}X
 \left[ \prod_\sigma \int_0^\infty\frac{-2\dif e_\gamma}{(ie_\gamma)^{3/2}}
 \right]
 \exp\left[
 -\frac{i}{2}\int\dif^2\sigma\left(
 \frac{-h}{e_\gamma}
 +e_\gamma
 \right)
 \right]
 ,
 \label{Schild-Mink_pos}
\end{align}
where $h$ is now the determinant of the Minkowskian induced metric:
\begin{equation}
 h=\det_{a,b}(\partial_aX^\mu\partial_bX_\mu)=\frac{1}{2}\left\{
  (\varepsilon^{ab}\partial_aX^i\partial_bX^j)^2
  -2(\varepsilon^{ab}\partial_aX^0\partial_bX^i)^2
 \right\}.
\end{equation}
Note that the integration measure for $X^D$ is also rotated
and the measure $\mathcal{D}X$ now contains that for $X^0$.

The Wick rotation in the opposite direction,
\begin{equation}
 X^D=-e^{i\theta}X^0
 ,
 \qquad
 e_g=-e^{i\theta}e_\gamma
 ,
 \label{Wick-rot-Schild-2}
\end{equation}
where $\theta$ varies from 0 to $-\pi/2$,
keeps the real part of the exponent negative definite as well 
if $e_\gamma\in (-\infty,0)$.
Note the minus sign in the rotation of $X^D$ is 
not relevant in the bosonic-string case
but it is relevant in the supersymmetric case
for making the action be in the standard form after the rotation.
We thus have
\begin{align}
 &\int\mathcal{D}X
 \left[ \prod_\sigma \int_0^\infty\frac{-\dif e_\gamma}{(ie_\gamma)^{3/2}}
 \right]
 \exp\left[
 -\frac{i}{2}\int\dif^2\sigma\left(
 \frac{-h}{e_\gamma}
 +e_\gamma
 \right)
 \right]
 \nonumber \\
 &=
 \frac{1}{2}Z=
 \int\mathcal{D}X
 \left[ \prod_\sigma \int_{-\infty}^0\frac{-\dif e_\gamma}{(ie_\gamma)^{3/2}}
 \right]
 \exp\left[
 -\frac{i}{2}\int\dif^2\sigma\left(
 \frac{-h}{e_\gamma}
 +e_\gamma
 \right)
 \right]
 ,
\end{align}
using Eq.~(\ref{Schild-Mink_pos}),
and the path integral is rewritten as
\begin{align}
 Z=\int\mathcal{D}X
 \left[ \prod_\sigma \int_{-\infty}^\infty\frac{-\dif e_\gamma}{(ie_\gamma)^{3/2}}
 \right]
 \exp\left[
 -\frac{i}{2}\int\dif^2\sigma\left(
 \frac{-h}{e_\gamma}
 +e_\gamma
 \right)
 \right]
 .
 \label{Schild-Mink}
\end{align}
Hence, this path-integral formulation for the Minkowskian Schild-type action
is equivalent to that for the Euclidean Schild-type action.

\begin{figure}[htbp]
 \centering
 \includegraphics[scale=0.5]{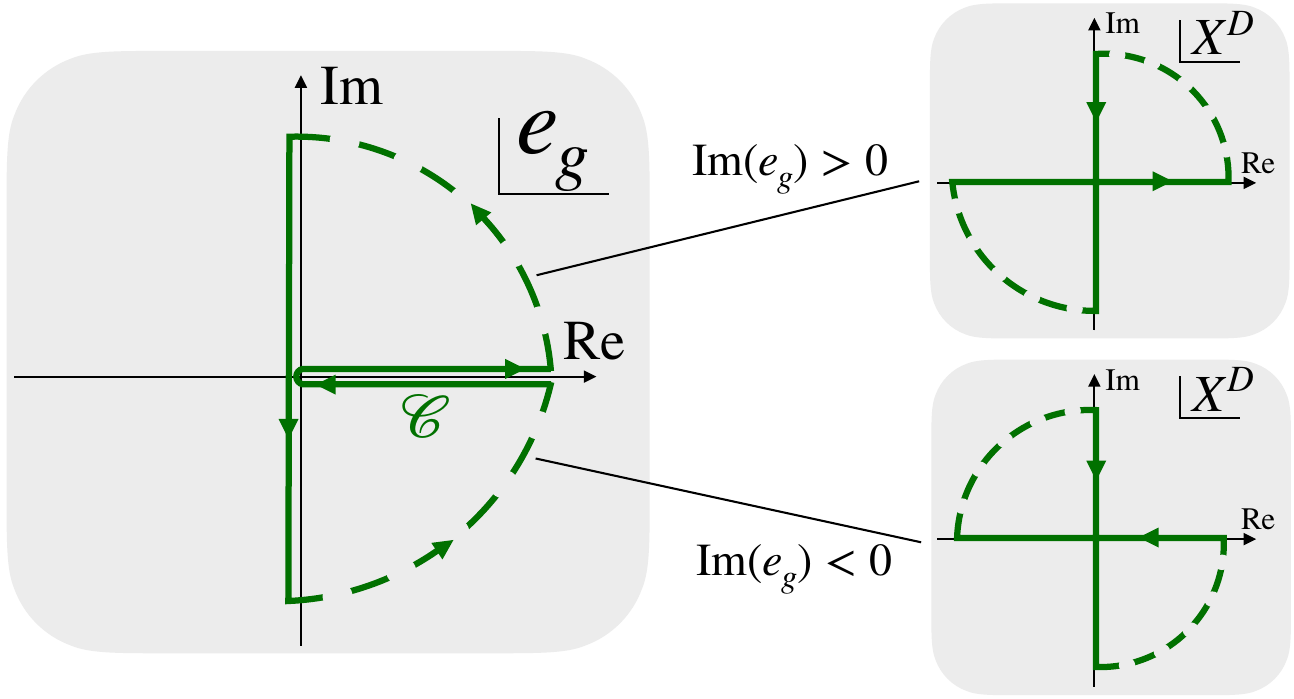}
 \caption{The contour of integration over $e_g$ and $X^D$.
  $\mathscr{C}$ represents the twofold contour on the real axis, coming from $+\infty-i0$ to $-0$ and then going to $+\infty+i0$. On the dashed arcs in the complex $e_g$ plane, $X^D$ is rotated differently depending on the sign of $\operatorname{Im}(e_g)$.}
 \label{fig:XD-eg-contour}
\end{figure}

Note that 
one can also show the equivalence between (\ref{Schild-Euc}) and (\ref{Schild-Mink}) 
by applying Cauchy's integral theorem to a closed contour 
that combines two Wick rotations (\ref{Wick-rot-Schild-1}) and (\ref{Wick-rot-Schild-2}) together, 
depicted in Fig.~\ref{fig:XD-eg-contour}.
More precisely, the closed contour for $X^D$ and $e_g$
is the limit of a union of the following six areas as $R\to +\infty$\,:
\begin{equation*}
 (X^D,e_g)=\left\{
 \begin{array}{l}
  (\zeta R\cos\varphi,\zeta R\sin\varphi+i0)\\
  (e^{\frac{\pi i}{2}\zeta}R\cos\varphi,e^{\frac{\pi i}{2}\zeta}R\sin\varphi)\\
  (i\zeta R\cos\varphi,i\zeta R\sin\varphi)\\
  (i\zeta R\cos(\varphi-\pi),i\zeta R\sin(\varphi-\pi))\\
  (-e^{-\frac{\pi i}{2}\zeta}R\cos(\varphi-\pi),-e^{-\frac{\pi i}{2}\zeta}R\sin(\varphi-\pi))\\
  (-\zeta R\cos(\varphi-\pi),-\zeta R\sin(\varphi-\pi)-i0)
 \end{array}
 \right.
\end{equation*}
with $\zeta\in[0,1]$ and $\varphi\in[0,\pi]$.

One can then show 
the equivalence to the path-integral formulation for the Minkowskian Polyakov-type action
by changing variables via $e_\gamma=\Lambda^0h_{11}$
and inserting
\begin{align}
 1=\left[ \prod_\sigma\int_{-\infty}^\infty\frac{\dif\Lambda^1}{(i\Lambda^0/h_{11})^{1/2}} \right]
 \exp\left[
 \frac{i}{2}\int\dif^2\sigma
 \frac{h_{11}}{\Lambda^0}\left( \Lambda^1-\frac{h_{01}}{h_{11}} \right)^2
 \right]
\end{align}
into the path integral.
We arrive at
\begin{align}
 Z=\int\mathcal{D}X
 \left[ \prod_\sigma \int_{-\infty}^\infty\frac{\dif\Lambda^0}{(\Lambda^0)^{2}}
 \int_{-\infty}^\infty\dif\Lambda^1
 \right]
 \exp\left[
 -\frac{i}{2}\int\dif^2\sigma
 \sqrt{-\gamma} \gamma^{ab}h_{ab}
 \right]
 ,
 \label{Polyakov-Mink}
\end{align}
where
\begin{align}
 \gamma^{ab}=e^{-\phi}
 \begin{pmatrix}
  -\frac{1}{\Lambda^0}&\frac{\Lambda^1}{\Lambda^0}\\
  \frac{\Lambda^1}{\Lambda^0}&\frac{-(\Lambda^1)^2+(\Lambda^0)^2}{\Lambda^0}
 \end{pmatrix}
 .
\end{align}
Again, the measure of path integration
$\frac{\dif\Lambda^0\dif\Lambda^1}{(\Lambda^0)^2}$
is reparametrisation invariant,
and one can freely add integration over the conformal factor, 
as long as one considers critical string theory,
to form the measure for the world-sheet metric
\begin{equation}
 \mathcal{D}\gamma
 =\mathcal{D}\phi\,\prod_\sigma\frac{e^\phi\dif\Lambda^0\dif\Lambda^1}{(\Lambda^0)^2}
 .
\end{equation}


\subsection{Polyakov to Nambu-Goto in the Minkowskian signature}

Let us now show that the path integral has a Nambu-Goto-type formulation 
equivalent to Polyakov's Euclidean path integral.
Again, the following arguments and proofs hold true 
even if one inserts operators to the path integral.

We first notice that the equivalent Minkowskian Polyakov-type formulation (\ref{Polyakov-Mink})
implicitly has $i\epsilon$ regulator terms as
\begin{align}
 Z=\lim_{\epsilon,\tilde\epsilon\to +0}\int\mathcal{D}X
 \left[ \prod_\sigma \int_{-\infty}^\infty\frac{\dif\Lambda^0}{(\Lambda^0)^{2}}
 \int_{-\infty}^\infty\dif\Lambda^1
 \right]
 \exp\left[
 -\frac{i}{2}\int\dif^2\sigma
 \left(
 \sqrt{-\gamma} \gamma^{ab}h_{ab}
 -i\epsilon|\Lambda^0|
 -i\frac{\tilde\epsilon (\Lambda^1-c)^2}{|\Lambda^0|}
 \right)
 \right]
 ,
 \label{Polyakov-Mink-reg}
\end{align}
because it was obtained by deformation of the contour through the Schild-type formulation.
Here, $\epsilon$, $\tilde\epsilon$ and $c$ are independent of $\Lambda^a$
but can depend on $X^\mu$ so that the terms are gauge-invariant.
These regulator terms can be understood as the result of implementation of the constraints associated with the reparametrisation symmetry
since the delta function for a constraint $\chi$ can be written as
\begin{equation*}
 \delta(\chi)
 =\lim_{\epsilon\to +0}\int_{-\infty}^\infty\frac{\dif\Lambda}{2\pi} e^{i\Lambda\chi-\epsilon|\Lambda|}
 =\lim_{\tilde\epsilon\to +0}\int_{-\infty}^\infty\frac{\dif\Lambda}{2\pi} e^{i\Lambda\chi-\tilde\epsilon\Lambda^2}
 .
\end{equation*}
The regulators are also naturally introduced 
in the Minkowskian Polyakov-type theory
because we are interested in scattering amplitudes,
which are expressed as the expectation values of 
incoming and outgoing operators 
in the ground state of perturbative string theory,
and thus the $i\epsilon$ prescription should be required
for the ground state.

After integrating out $\Lambda^1$ and then $\Lambda^0$,
we obtain
\begin{align}
 Z&=\lim_{\epsilon',\tilde\epsilon'\to +0}\int\mathcal{D}X
 \left[ \prod_\sigma \int_{-\infty}^\infty\frac{-\dif e_\gamma}{(ie_\gamma)^{3/2}}
 \right]
 \exp\left[
 -\frac{i}{2}\int\dif^2\sigma\left(
 \frac{-h}{e_\gamma}
 +e_\gamma
 -i\epsilon' |e_\gamma|
 -i\frac{\tilde\epsilon'}{|e_\gamma|}
 \right)
 \right]
  \nonumber \\
 &=\lim_{\epsilon''\to +0}\int\mathcal{D}X \,
 \prod_\sigma
 \left(
 \frac{\exp\left[
 -i\iDelta\Sigma\sqrt{-h-i\epsilon''}
 \right]}{\sqrt{-h-i\epsilon''}}
 +\frac{\exp\left[
 i\iDelta\Sigma\sqrt{-h+i\epsilon''}
 \right]}{\sqrt{-h+i\epsilon''}}
 \right)
 ,
 \label{Nambu-Goto-Mink-wo-s}
\end{align}
where $\iDelta\Sigma$ is the infinitesimal world-sheet area element.
In order to make the integral in a suitable form for path integration,
we introduce a sign function $s(\sigma)=\pm 1$ and rewrite it as
\begin{align}
 Z=\lim_{\epsilon\to +0}\int\mathcal{D}X 
 \left[ \prod_\sigma\sum_{s(\sigma)=\pm 1}
 \frac{1}{\sqrt{-h-i\epsilon s}}
 \right]
 \exp\left[
 -i\int\dif^2\sigma\, s\sqrt{-h-i\epsilon s}
 \right]
 .
 \label{Nambu-Goto-Mink}
\end{align}
This is the Nambu-Goto-type formulation equivalent to Polyakov's Euclidean path integral.
Therefore,
the Polyakov, Schild and Nambu-Goto types in the Euclidean and Minkowskian signatures
are all quantum-mechanically equivalent in this sense.

A remarkable property is readily seen in Eq.~(\ref{Nambu-Goto-Mink-wo-s}).
This integral is non-zero in general when $h<0$ at all points in the world-sheet
while it becomes zero when there is a region where $h>0$
because of the cancellation of $s=1$ and $s=-1$ contributions.
We interpret this as ``stringy causality''---any space-like propagation of a string
is prohibited in path integration.
If we see the contribution with $s=1$ as that of the fundamental string,
we take the contribution with $s=-1$ as that of the anti-fundamental string
because the coupling to the Kalb-Ramond $B_{\mu\nu}$ field in the world-sheet action,
if exists, does not contain $s$ so that 
the opposite sign of $s$ should represent a negative NS-NS charge.
Therefore, this type of causality is realised
thanks to the existence of the anti-fundamental string with the opposite overall sign in the action---this mechanism resembles the realisation of causality in standard quantum field theory to some extent.

Physical interpretation of such an anti-F-string, 
which looks travelling backward in time,
has not been fully comprehended yet.
However, there are some notions about the interpretation:
\begin{itemize}
 \item The opposite overall sign would not be problematic because we may say 
 there is no world-sheet time direction in the first place,
 that is to say, 
 quantum time evolution of a perturbative string is trivial 
 as the Hamiltonian contains only terms with the Lagrange multipliers $\Lambda^a$
 and so vanishes when the constraints are imposed.
 \item The anti-F-string with $\sqrt{-\gamma}\gamma^{00}=-1/\Lambda^0>0$,
 which is in the ``wrong sign'',
 can be reinterpreted as an F-string 
 with the ``right sign'', $\sqrt{-\gamma}\gamma^{00}<0$,
 by an exchange of $\sigma^0$ and $\sigma^1$.
 Namely, this unusual anti-F-string should be nothing but 
 the counterpart of an F-string in open-closed string duality.
 \item The opposite sign in the action indicates
 it could be related to the ghost D-brane \cite{Okuda:2006fb}.
\end{itemize}

\subsection{Generalisations}
In the case of critical type II superstring,
we utilise the Green-Schwarz formalism 
to show the equivalences for path-integral formulations
(See Ref.~\cite{Asano:2024edo} for details).
The fermionic fields $\theta^1$ and $\theta^2$,
which are the superpartners of $X^\mu$,
have redundancy associated with fermionic gauge symmetry,
known as $\kappa$ symmetry.
For a proof of the equivalences,
we fix the $\kappa$ symmetry\footnote{
One needs to rotate $\theta^2$ in the path integral as $\theta^2\to i\theta^2$
in order to use these gauges.
}
by $\theta^2=i\theta^1$ 
for the type IIB string
and $\theta^2=\bar\theta^{1 \dagger}=i\Gamma^0\theta^{1}$
for the type IIA string.
Then, the induced metric in the gauge is written as
\begin{align}
 h_{ab}=\Pi_a^i\Pi_b^i-\partial_aX^0\partial_bX^0,
\end{align}
embedded in the Minkowski target spacetime,
where $\Pi_a^\mu=\partial_aX^\mu-i(\bar\theta^{1}\Gamma^\mu\partial_a\theta^1+\bar\theta^{2}\Gamma^\mu\partial_a\theta^2)$.
Note $\Pi_a^i=\partial_aX^i$ for the gauge-fixed type IIB theory.
Thus the determinant is
\begin{align}
 h=\frac{1}{2}\left\{ (\varepsilon^{ab}\Pi_a^i\Pi_b^j)^2-2(\varepsilon^{ab}\partial_aX^0\Pi_b^i)^2 \right\}
 ,
 \label{det-h}
\end{align}
which is able to be Wick-rotated (back) in the same way as Eq.~(\ref{exponent-Wick-rot}).
Since the Minkowskian path integral in the Green-Schwarz formalism is in the following form,
\begin{align}
 Z=\int\mathcal{D}X
 \mathcal{D}\theta
 \mathcal{D}\gamma\,
 \exp\left[
 -\frac{i}{2}\int\dif^2\sigma
 \sqrt{-\gamma} \gamma^{ab}h_{ab}
 \right]
 \exp\left[ i \iDelta S_{\rm f}[X,\theta] \right]
 ,
\end{align}
where $\iDelta S_{\rm f}[X,\theta]$, the additional Wess-Zumino-type action for $\kappa$ symmetry,
is separated from the integration over the world-sheet metric $\gamma_{ab}$,
it is clear that the proof of the equivalences in the previous subsections 
is applicable to the supersymmetric case for the flat target space.

It is possible to generalise the proof to curved target space.
It is obvious from the structure of $h$, seen in Eq.~(\ref{det-h}),
that Cauchy's integral theorem with the Wick rotation in Eq.~(\ref{exponent-Wick-rot}) is applicable 
if the target space metric $G_{\mu\nu}(X)$ is
static ($\pder{G_{\mu\nu}}{X^0}=0$ and $G_{0i}=0$).
Therefore,
the equivalences for the path-integral formulations should be valid 
for a general static background 
if the string theory is free from the conformal anomaly.

\section{Path integral for the matrix model}

The matrix model is obtained by matrix regularisation of string perturbation theory,
by which a function on a compact base space is mapped to a set of matrices.
Since a set of matrices can describe multiple geometries with arbitrary topology
as non-commutative geometry,
we expect that the matrix model describes
multi-body systems of interacting superstrings non-perturbatively.

For the IKKT-model action,
matrix regularisation is applied to a gauge-fixed Schild-type action
of type IIB string theory.
However, there is an ambiguity in applying matrix regularisation
as the resultant matrix-model action may depend on
whether the world-sheet theory is Euclidean or Minkowskian,
which gauge symmetry is fixed, and so on.
In this paper, we consider matrix regularisation applied
to the Minkowskian Schild-type action without Wick rotation 
with only $\kappa$ symmetry fixed.
Though it may look invalid at first sight
since the target space is Lorentzian,
one can apply the matrix regularisation for the following reason.
First, the world-sheet coordinates are just parameters in the Schild-type action,
which contains no metric for the base space.
In addition, as we are interested in scattering amplitudes,
the world-sheet should have punctures in actual settings,
and thus it admits Lorentzian induced metric even if the base space is compact.
Therefore, one can matrix-regularise the Minkowskian Schild-type action 
by considering a compact base space with punctures.
Note that, because of the fuzziness of matrix-regularised geometry,
the information of punctures will be washed away upon the regularisation
and the information of incoming and outgoing string states
should be purely encoded by vertex operators.

After gauge-fixing of $\kappa$ symmetry by $\theta^2=i\theta^1$,
the Schild-type action becomes 
\begin{align}
 S_{\rm Schild}=\int\dif^2\sigma\left[
 \frac{1}{4e_\gamma} \{X^\mu,X^\nu\}^2
 +2i\psi^T\Gamma_\mu\{X^\mu,\psi\}
 -\frac{e_\gamma}{2}
 \right]
 ,
\end{align}
where $\psi=\theta^1-i\theta^2=2\theta^1$
and the Poisson bracket is defined by
$\{f,g\}=\varepsilon^{ab}\partial_af\partial_bg$.
By matrix regularisation \cite{Hoppe82mr}
$X^\mu(\sigma)\mapsto X^\mu$, $\psi(\sigma)\mapsto\psi$, $e_\gamma(\sigma)\mapsto -Y$
with
\begin{equation*}
 \{ \cdot,\cdot \} \mapsto \frac{N}{i}[\cdot,\cdot ],
 \qquad
 \frac{1}{\pi}\int\dif^2\sigma\mapsto\frac{1}{N}\tr
 ,
\end{equation*}
without Wick rotation,
we obtain the NBI-type IKKT matrix model
\begin{align}
 S_{\rm NBI}
 =N\tr\left(
 \frac{1}{4}Y^{-1}[X^\mu,X^\nu]^2
 +\frac{1}{2}\psi^T\Gamma_\mu[X^\mu,\psi]
 +Y+\frac{i}{N}(N+\tfrac{1}{2})\ln(-iY)
 \right)
 .
 \label{NBI-action}
\end{align}
Note that the final term is absent in the original IKKT model
because $e_\gamma$ was not matrix-regularised in the original paper~\cite{Ishibashi:1996xs}.
Here, the final term comes from the integration measure
but we choose its coefficient 
so that the matrix model holds ``causality'' property.
Unlike the Euclidean version of this matrix model, 
proposed in Ref.~\cite{Fayyazuddin:1997yf},
we take $Y$ as a Hermitian matrix without constraining it to be positive definite.

The Minkowskian NBI-type IKKT model also has (dynamical) supersymmetry
in the large-$N$ limit\footnote{
For finite $N$, actually, the real part of the action is invariant
under a supersymmetry transformation with
\begin{equation*}
 \delta^{(1)} Y
 =i\left\{
 Y(
 -\tfrac{1}{4}[X^\rho,X^\sigma]^2+Y^2
 )^{-1}Y \,,\;
 \left\{ \epsilon^{(1)T}\Gamma_{\mu\nu\lambda}\psi, \{[X^\nu,X^\lambda],[X^\mu,Y^{-1}]\}
 \right\}
 \right\}
 ,
\end{equation*}
which is a counterpart of $\kappa$ transformation in perturbative string theory \cite{Asano:2024edo}.
} \cite{Fayyazuddin:1997yf}
\begin{equation}
 \delta^{(1)} X^\mu=i\epsilon^{(1)T}\Gamma^\mu\psi
 ,
 \qquad
 \delta^{(1)} \psi
 =
 -\frac{i}{4}\{Y^{-1},[X_\mu,X_\nu]\}\Gamma^{\mu\nu}
 \epsilon^{(1)}
 ,
\end{equation}
when $\delta^{(1)} Y=0$,
where $\epsilon^{(1)}$ is a spinor parameter of the supersymmetry transformation.
In addition, the kinematical supersymmetry 
$\delta^{(2)} X^\mu=\delta^{(2)} Y=0$,
$\delta^{(2)} \psi=\epsilon^{(2)}$
is preserved
when the matrices $X^\mu$ and $\psi$ each have a trace mode.

Let us observe the causality structure of the NBI-type IKKT model.
We first set the path-integration measure for the matrices to
the standard $SU(N)$-invariant measure.
By integrating out $Y$, which is Itzykson-Zuber-type integration \cite{Itzykson:1979fi,Mehta:1981xt}, 
using the same integral formula in Eq.~(\ref{Nambu-Goto-Mink-wo-s}),
we obtain
\begin{align}
 &\lim_{\epsilon,\tilde\epsilon\to +0}\int 
 \mathcal{D}Y\,
 \exp\left[
 iN\tr \bigg( 
 \frac{1}{4}Y^{-1} M
 +Y
 +\frac{i}{N}\left( N+\frac{1}{2} \right) \ln (-iY)
 +i\epsilon Y^2
 +i\tilde\epsilon Y^{-2}
 \bigg)
 \right]
 \nonumber \\
 &\propto \lim_{\epsilon,\tilde\epsilon\to +0}
 \int\frac{[\dif y]\Delta(y)^2}{\Delta(y^{-1})\Delta(m)}
 \exp\left[
 iN\sum_{i=1}^N \bigg( 
 \frac{1}{4}y_i^{-1} m_i
 +y_i
 +\frac{i}{N}\left( N+\frac{1}{2} \right) \ln (-iy_i)
 +i\epsilon y_i^2
 +i\tilde\epsilon y_i^{-2}
 \bigg)
 \right]
 \nonumber \\
 &=\lim_{\epsilon'\to +0}
 \left( -\sqrt{\frac{\pi}{N}} \right)^{N}
 \Delta(m)^{-1}\det_{i,j}
 \left[
 \left( \frac{1}{iN}\pder{}{\alpha}\right) ^{j-1}
 \left(
 \frac{e^{-iN\sqrt{m_i-i\epsilon'}\sqrt{\alpha}}}{\sqrt{m_i-i\epsilon'}}
 +\frac{e^{iN\sqrt{m_i+i\epsilon'}\sqrt{\alpha}}}{\sqrt{m_i+i\epsilon'}}
 \right)
 \right]_{\alpha\to 1}
 ,
 \label{Y-integration}
\end{align}
where $M=[X^\mu,X^\nu]^2$,
$m_i$ denotes the $i$th eigenvalue of $M$,
and $\Delta(m)$ is the Vandermonde determinant for $m_i$.
The resultant determinant is zero if at least one eigenvalue of $M$ is negative;
the way it vanishes is exactly similar to the cancellation in perturbative string theory,
which we observed in section~\ref{sec:EtoM}.
Eq.~(\ref{Y-integration}) is further computed to be
\begin{align}
  &=\lim_{\epsilon\to +0}
   \left( -\sqrt{\tfrac{\pi}{N}} \right)^{N}
   \Delta(m)^{-1}\det_{i,j}
   \left[
   \sum_{s=\pm 1}
   -s\left( -s\frac{\sqrt{m_i-i\epsilon s}}{2} \right) ^{j-2}
   \frac{e^{-isN\sqrt{m_i-i\epsilon s}}}{2}
   \right]
   \nonumber \\
  &=\lim_{\epsilon\to +0}
   \left( -\sqrt{\tfrac{\pi}{N}} \right)^{N}
   \left[
   \prod_{k=1}^N\sum_{s_k=\pm 1}
   \right]
   \frac{(-2)^{-\frac{N(N-1)}{2}}}{\prod_{i,j<i}(s_i\sqrt{m_i-i\epsilon s_i}+s_j\sqrt{m_j-i\epsilon s_j})}
   \prod_{i=1}^N
   \left(
   \frac{e^{-is_iN\sqrt{m_i-i\epsilon s_i}}}{\sqrt{m_i-i\epsilon s_i}}
   \right)
   ,
   \label{I_M}
\end{align}
where the first equality in Eq.~(\ref{I_M}) is obtained
by adding scalar multiples of the $k$th column of the matrix for the determinant
with $k<j$ to the $j$th column 
so that the terms with $\sqrt{m_i}$ to the power of lower than $(j-2)$ 
are cancelled,
and the second equality is obtained
by putting the summations on $s$ outside of the determinant
and using $\Delta(s_i\sqrt{m})/\Delta(m)=\prod_{i,j<i}(s_i\sqrt{m_i}+s_j\sqrt{m_j})^{-1}$.
It is evident in this expression that the causality structure is almost identical to
that in perturbative string theory.

We may interpret the matrix $([X^\mu,X^\nu]^2)^{\frac{1}{2}}$ 
as ``fuzzy world-sheet'' because 
$\tr ([X^\mu,X^\nu]^2)^{\frac{1}{2}}$ approaches 
a time-like area $\int\dif^2\sigma\sqrt{-h}$
in the large-$N$ limit,
by a similar discussion in Ref.~\cite{Fayyazuddin:1997yf}.
We then view fuzzy world-sheet with 
all the eigenvalues of $[X^\mu,X^\nu]^2$ positive as time-like
and see from the causality structure that
only the time-like fuzzy world-sheet contributes to the path integral.
Therefore, 
the Minkowskian NBI-type IKKT model can be interpreted as a causal matrix model.

Before closing the section,
we note that the Minkowskian NBI-type IKKT model still needs
an additional regularisation prescription 
other than the terms $i\epsilon Y^2+i\tilde\epsilon Y^{-2}$ 
in order to define a finite path integral
by dealing with divergence coming from the Lorentz invariance.
This is because the measure of integration contains 
infinite volumes of the Lorentz group
and one way to overcome the divergence is 
to ``gauge-fix'' the Lorentz symmetry by introducing a Faddeev-Popov determinant
\cite{Asano:2024def}.
Nonetheless, it may also be possible that the divergence disappears 
without gauge-fixing of the Lorentz symmetry
but instead by fixing $SU(N)$ symmetry and 
taking the large-$N$ limit, 
because 
perturbative string theory with a fixed genus 
has no divergence originating from the Lorentz symmetry 
after fixing the world-sheet diffeomorphism.

\section{Summary and discussion}
The Euclidean perturbative string theory is 
quantum-mechanically equivalent to its Minkowskian version 
in terms of path integration.
We showed the quantum-mechanical equivalence among 
the Polyakov, Schild and Nambu-Goto-type formulations 
in the case of critical (bosonic and type II) string theory,
by considering the appropriate measure of path integration.

In particular, the Minkowskian path-integral formulation of the Nambu-Goto type 
equivalent to Polyakov's Euclidean path integral has
anti-F-string contribution explicitly,
and it is responsible for the stringy causality,
by which string propagation with a space-like area ($\det h_{ab}>0$) is prohibited.
Remarkably, the causality structure is already implemented
in the standard perturbative string theory,
but hidden in the Euclidean path-integral formulation.
To fully understand the stringy causality,
we need to verify its exact relationship with 
the causality in standard quantum field theory\footnote{
The cancellation in the stringy causality might be related to
the mechanism of causality in the loop-tree duality formulation \cite{deJesusAguilera-Verdugo:2021mvg}.
It would be also interesting to study the relation with
the world-line formalism of quantum field theory (See for example Ref.~\cite{Schubert:2001he} for review).
}.

By applying matrix regularisation 
to the Minkowskian Schild-type path-integral formulation
of the type IIB string,
the Minkowskian NBI-type IKKT model (\ref{NBI-action})
is obtained as a causal matrix model.
We interpreted the matrix model as theory on fuzzy world-sheet,
and just like the Minkowskian perturbative string theory,
we observed that only the fuzzy world-sheet that corresponds to a time-like area 
contributes to the matrix-model path integral.
Since the stringy causality is realised 
thanks to the anti-F-string contribution, which may be related to a ghost D-string \cite{Okuda:2006fb},
it would be interesting to study the relationship with
the matrix model obtained by ghost D-instantons \cite{Terashima:2006qm}.

Note also that the logarithmic potential is analogous \cite{Kristjansen:1997mx} to the Penner model \cite{Penner:1988cza,Distler:1990mt},
which can generate Euler characteristic of world-sheet moduli space.
Hence, although this matrix model may belong to the same universality class as the original IKKT matrix model \cite{Fukuma:1997en},
the Minkowskian NBI-type IKKT matrix model may allow us to reveal much clearer interpretation of the matrices as degrees of freedom of the superstring and hence as gravitational degrees of freedom.

\section*{Acknowledgement}
The author thanks Masafumi Fukuma, Denjoe O'Connor and Shigeki Sugimoto for valuable discussions.
This work was supported by JSPS KAKENHI Grant Number JP24K07036.



\bibliographystyle{JHEPnote}
\bibliography{ikktq-corfu}

\end{document}